\documentclass[conference]{IEEEtran}
\usepackage{graphicx}
\usepackage{amssymb}
\usepackage{amsmath}
\usepackage{calc}
\usepackage{array}
\usepackage{subfigure}
\usepackage{graphicx}
\usepackage{algorithm}
\usepackage{algorithmic}

\newtheorem{remark}{Remark}

\begin{document}
\title{\huge Grouping Based Blind Interference Alignment for $K$-user MISO Interference Channels}

\author{Heecheol Yang, Wonjae Shin, and Jungwoo Lee \\
Department of Electrical and Computer Engineering, Seoul National University, Seoul, Korea \\
E-mail: hee2070@wspl.snu.ac.kr, \{wonjae.shin, junglee\}@snu.ac.kr \vspace{-3mm}
}

\maketitle

\begin{abstract}
We propose a blind interference alignment (BIA) through staggered antenna switching scheme with no ideal channel assumption.
Contrary to the ideal assumption that channels remain constant during BIA symbol extension period,
when the coherence time of the channel is relatively short, channel coefficients may change during a given symbol extension
period.
To perform BIA perfectly with realistic channel assumption, we propose a grouping based supersymbol structure
for $K$-user interference channels which can adjust a supersymbol length to given coherence time.
It is proved that the supersymbol length could be reduced significantly by an appropriate grouping.
Furthermore, it is also shown that the grouping based supersymbol achieves higher degrees of freedom than the
conventional method with given coherence time.
\end{abstract}

\begin{keywords}
Blind interference alignment, degrees of freedom, supersymbol, coherence time.
\end{keywords}

\section{Introduction}
\label{sec_intro}
Interference is a limiting factor in wireless networks although the demand for high data rate is ever more increasing.
To characterize the performance of interference-limited networks, degrees of freedom (DoF) has been widely
used. DoF represents the number of independent signal dimensions that are not disrupted by interference signals.
The conventional approach to mitigate interference, which exploits the orthogonality between signals, has no merit
within the framework of DoF.
In recent years, interference alignment (IA) has attracted significant attention \cite{IA}-\cite{XIA}.
By using IA concept, each user achieves 1/2 DoF for $K$-user interference channels (IC), moreover, a total of $K/2$ DoF
could be achieved although the number of users increases.
However, there is a great impediment for IA to be implemented in real communication systems.
The IA technique exploits global channel state information at transmitter (CSIT) to align interference signals
in small subspaces and guarantee separate subspaces to the desired signal.
The necessity for global CSIT imposes much burden on communication systems since it causes many problems related to
the channel state feedback.

According to the result of \cite{Huang}-\cite{Vaze}, interference network cannot achieve more than 1 DoF in the absence of
CSIT. Recently, it was reported that IA technique can be performed with no CSIT when
the channel state satisfies certain required assumptions with regard to coherence time/bandwidth \cite{blind}-\cite{blind2},
which is called \emph{blind interference alignment} (BIA).
It was demonstrated that a total of $\frac{MK}{M+K-1}$ DoF is achieved for $K$-user $M \times 1$
multiple input single output (MISO) broadcast channel with no CSIT.
To make BIA more practical, staggered antenna switching scheme was proposed \cite{Gou}.
In \cite{Gou}, each receiver is equipped with a reconfigurable antenna which can select different preset modes to control
the pattern of the channel state variation as needed.
This concept was applied to the $K$-user MISO IC in \cite{IC}-\cite{IC2}.
For the IC setting that each user has different number of preset modes,
they are attuned to transmit signal vectors
depending on the number of preset modes to construct an appropriate supersymbol.

The great interest of this paper is to propose a grouping based supersymbol structure within a given supersymbol length
for $K$-user MISO IC. Although BIA scheme can exploit the benefit of staggered antenna switching,
it needs a nontrivial channel assumption. That is the coherence time of the channel is long enough, thereby channel
coefficients remain constant across a supersymbol. If the channel state varies,
interference signals are not clearly removed through BIA. Therefore, it is important for the supersymbol to have a short
symbol extension period since we cannot control the coherence time of the channel which is the function of Doppler spread
in most communication scenarios. In this paper, the channel state is assumed to be constant during limited block length.
This assumption could be more effective because IA schemes which require CSIT are difficult to be realized
when the channel state varies relatively fast.
Accordingly, we propose a grouping based supersymbol structure for $K$-user MISO IC
which aligns interference signals within relatively short symbol extension period.
It can adjust the supersymbol length in accordance with the coherence time
of the channel by changing the number of groups for users and preset modes.
Therefore, the proposed scheme aims to achieve the maximum DoF within limited supersymbol length.

\emph{Notation:}
For a vector $\mathbf{a}$, $\| \mathbf{a} \|$ means Euclidean norm of $\mathbf{a}$.
For a matrix $\mathbf{A}$, $\mathbf{A}^{T}$ means transpose of $\mathbf{A}$.
$\mathbb{E}[\cdot]$ represents an expected value. $\mathbf{0}_{m\times n}$ is a $m \times n$ zero matrix.
$\mathcal{O}(\cdot)$ describes the limiting behavior of a function when the argument tends toward infinity.


\section{System Model}
\label{sec_model}
\begin{figure}[t]
    \centerline{\includegraphics[width=8.0cm]{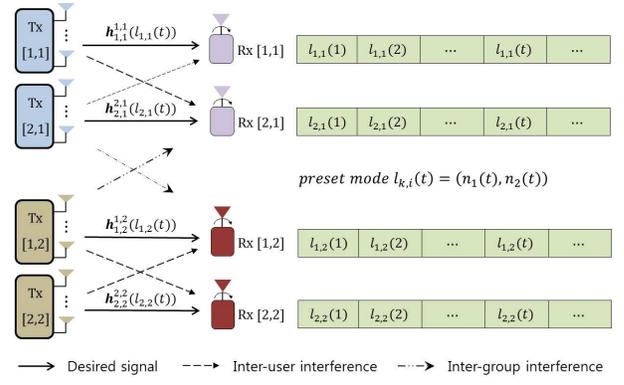}}
    \caption{$K$-user MISO IC with BIA when $(K,K_{G})=(4,2)$.}
    \label{system_model}
    \vspace{-3mm}
\end{figure}

Consider a system model for the $K$-user MISO IC BIA scheme with reconfigurable antenna switching.
The system contains $K$ transmitters with $M_{k}$ transmit antennas
and $K$ receivers equipped with single reconfigurable antenna that can switch among $M_{k}$ preset modes.
Each transmitter serves its own paired receiver. The transmitted signal from the other transmitter acts as a interference to the receiver.
We divide $K$ users into $K_{G}$ groups when $K$ is divisible by $K_{G}$,
thereby $k^{th}$ user in the group $i$ is denoted as user $[k,i]$
where $k \in \{1,2,\ldots,K_{E}\}$, $i \in \{1,2,\ldots,K_{G}\}$, and $K_{E}=K/K_{G}$.
After users are grouped, interference signals are classified for convenience.
The interference signal from intra-group transmitter is called inter-user interference (IUI) and that from
inter-group transmitter is called inter-group interference (IGI) from now on.
The number of preset modes for user $[k,i]$ is denoted as $M_{k,i}$.
We also partition $M_{k,i}$ preset modes into $M_{G_{i}}$ groups for group $i$ users
when $M_{1,i},\ldots,M_{K_{E},i}$ are divisible by $M_{G_{i}}$.
Then, $M_{k,i}/M_{G_{i}}=M_{E_{k,i}}$ preset modes belong to each preset mode group.
We define the $M_{E}$ set for group $i$ as $\{M_{E_{1,i}},\ldots,M_{E_{K_{E},i}}\}$.
The preset mode $(m_{1},m_{2})$ implies the $m_{1}^{th}$ mode in group $m_{2}$
where $m_{1} \in \{1,2,\ldots,M_{E_{k,i}}\} , m_{2} \in \{1,2,\ldots,M_{G_{i}}\}$.
Since receivers change their preset modes according to the time slot, we denote a preset mode of user $[k,i]$ at time $t$ as $l_{k,i}(t)$.
The channel matrix from the transmitter $[k',i']$ to the receiver $[k,i]$ at time $t$ is denoted by
$\mathbf{h}_{k,i}^{k',i'}(l_{k,i}(t)) \in \mathbb{C}^{1 \times M_{k',i'}}$.
It is assumed that channel coefficients are independent and identically distributed (i.i.d.)
so that any $M_{k',i'}$ of them for transmitter $[k',i']$ are linearly independent, almost surely.
The received signal for the user $[k,i]$ at time $t$ is
\begin{eqnarray}
y_{k,i}(t)=\sum\limits_{i'=1}^{K_{G}}\sum\limits_{k'=1}^{K_{E}}\mathbf{h}_{k,i}^{k',i'}(l_{k,i}(t))\mathbf{x}^{k',i'}(t)+
z_{k,i}(t),
\end{eqnarray}
where $\mathbf{x}^{k',i'}(t) \in \mathbb{C}^{M_{k',i'} \times 1}$ is a $M_{k',i'} \times 1$ transmitted signal
from the transmitter $[k',i']$, and $z_{k,i}(t)$ is the additive white Gaussian noise with $\mathcal{CN}(0,1)$.
Transmitters are subjected to the average transmit power constraint $\mathbb{E}[\| \mathbf{x}^{k',i'}(t) \|^{2}] \leq P$.
These are illustrated at Fig. \ref{system_model} for the case of $(K,K_{G})=(4,2)$.

Meanwhile, we call the predetermined sequence of antenna switching to construct supersymbol as preset mode pattern.
The preset mode pattern is determined by the set of each user's number of preset modes
to decide the number of transmit vectors appropriately.
To construct supersymbol structure as proposed,
the element $(m_{1})$ pattern and the group $(m_{2})$ pattern are designated, respectively.
The preset mode pattern of user $[k,i]$ is expressed as a Cartesian product
\footnote{In this paper, Cartesian product for sequences has a slightly different meaning from a conventional use for sets.
We define the operation of Cartesian product for two sequences
$A$ and $B$, denoted by $A \times B$, as a sequence of all ordered pairs $(a_{i},b_{j})$ such that $a_{i}$ and $b_{j}$ are $\textrm{i}^{th}$ element of $A$ and $\textrm{j}^{th}$ element of $B$, respectively.} as follows
\begin{eqnarray}
\mathbf{m}_{k,i} = (m_{1} \textrm{ pattern of user $[k,i]$}) \times \\ \nonumber
  (m_{2} \textrm{ pattern of user $[k,i]$}). \hspace{-5mm}
\end{eqnarray}
Suppose $\mathbf{m}_{k,i} = (p_{1},\ldots,p_{L_{1}})\times(q_{1},\ldots,q_{L_{2}})$
where $L_{1}$ and $L_{2}$ mean the lengths of $m_{1}$ and $m_{2}$ patterns, respectively,
$p_{1},\ldots,p_{L_{1}} \in \{1,\ldots,M_{E_{k,i}}\}$, and $q_{1},\ldots,q_{L_{2}} \in \{1,\ldots,M_{G_{i}}\}$.
$\mathbf{m}_{k,i}$ indicates that receiver $[k,i]$ has
a predetermined preset mode pattern as $((p_{1},q_{1}),\ldots,(p_{L_{1}},q_{1}),\ldots,(p_{1},q_{L_{2}}),\ldots,(p_{L_{1}},q_{L_{2}}))$ during $L(=L_{1}L_{2})$ symbol extension period.

There are some more assumptions for the system model.
To apply a more reasonable channel model, channel coefficients are assumed to remain constant
during an $L$-symbol extension period which is determined by the channel's coherence time.
Transmitters have no CSIT and receivers know perfect local channel state.
Lastly, the sum DoF is defined as the pre-log factor of the achievable sum rate. The individual DoF achieved by user $[k,i]$
and the sum DoF are expressed as
\begin{eqnarray}
d_{k,i}\triangleq \lim_{\textrm{SNR}\to\infty}\frac{R_{k,i}(\textrm{SNR})}{\textrm{log(SNR)}} \textrm{ and } \textrm{DoF}_{\textrm{sum}}=\sum\limits_{^{\forall} k,i}
 d_{k,i},
\end{eqnarray}
where $R_{k,i}(\textrm{SNR})$ denotes the achievable rate of user $[k,i]$ for the average power constraint $P$.

\section{Grouping based supersymbol structure}
\label{sec_grouping}
As mentioned in Section \ref{sec_intro}, interference signals are not aligned properly
when the channel state varies during the symbol extension period.
The supersymbol length for $K$-user MISO IC is
$\prod\nolimits_{k=1}^{K}(M_{k}-1)+\sum\nolimits_{k=1}^{K}\left(\prod\nolimits_{q=1,q\neq k}^{K}(M_{q}-1)\right)$,
when $K$ receivers are equipped with $M_{1},\ldots,M_{K}$ preset modes, respectively.
To reduce the supersymbol length, we propose a grouping based supersymbol structure which aims to align interference signals
with short symbol extension period.
The proposed successive alignment strategy aligns IUIs in a group by the $m_{1}$ pattern design and
aligns IGI by the $m_{2}$ pattern design sequentially.
This strategy drives a supersymbol length to decrease significantly.
When we consider only intra-group scenario to align IUI initially, the $m_{1}$ pattern for users in group $i$ is
determined as if there are $K_{E}$ users who have $M_{E_{1,i}},\ldots,M_{E_{K_{E},i}}$ preset modes, respectively.
Subsequently, after IUI is canceled, the $m_{2}$ pattern for each group is for $K_{G}$ groups
which have $M_{G_{1}},\ldots,M_{G_{K_{G}}}$ preset mode groups, respectively.
It is guaranteed that desired signals occupy separate subspaces from interference subspace by the grouping based supersymbol
structure when the condition on the number of preset modes are satisfied by appropriate grouping.
\vspace{1mm}
\begin{itemize}
	\item \textbf{Grouping based BIA condition: } \begin{eqnarray}M_{E_{k,1}}=M_{E_{k,2}}=\cdots=M_{E_{k,K_{G}}},
	\label{eq:con}
	\end{eqnarray}
		\end{itemize}
for $k \in \{1,2,\ldots,K_{E}\}$.
When the grouping result satisfies condition (\ref{eq:con}), each user group has the same $M_{E}$ set.
Moreover, receivers suffers IGI only from the transmitter in other group which has the same $m_{1}$ pattern
since IGI from the others are aligned due to the same supersymbol structure for IUI alignment between groups.
Since the supersymbol is determined by each user's number of preset modes, the $M_{E}$ set for each user group
needs to be the same to have a common supersymbol for IUI alignment.
If each group has a different $M_{E}$ set, IGI still remains although all the groups have an appropriate $m_{2}$ pattern.
As a result, the common $M_{E}$ set can be denoted as
\begin{eqnarray}
\{M_{E_{1,i}},\ldots,M_{E_{K_{E},i}}\}=\{M_{E_{1}},\ldots,M_{E_{K_{E}}}\},
\end{eqnarray}
for $i \in \{1,\ldots,K_{G}\}$.
After preset modes and users are grouped to satisfy condition (\ref{eq:con}),
the preset mode pattern is also determined according to the grouping based strategy.
Then, the preset mode pattern for user $[k,i]$ is
\begin{eqnarray}
\mathbf{m}_{k,i}=(m_{1} \textrm{ pattern for $k^{th}$ user in each group}) \times \\ \nonumber
(m_{2} \textrm{ pattern for users in group $i$}).
\label{eq:mki}
\end{eqnarray}
The transmit signal for each transmitter is determined according to the preset mode pattern.
Let us explain a simple example to show the grouping based supersymbol design easily.

Consider a simple case that four users have 6, 6, 4, and 4 preset modes, respectively, i.e., $\mathcal{M}=\{6,6,4,4\}$.
They are grouped in order that
two users in a group have 6, 4 preset modes.
We decide a representative preset mode group number for each user group as
$M_{G_{1}}=2, M_{G_{2}}=2$ to satisfy condition (\ref{eq:con}).
Then, the $M_{E}$ set for each group is the same as $\{6/2,4/2\}=\{3,2\}$.
After grouping users in this manner, the preset mode pattern can be determined according to the grouping result.
The $m_{1}$ pattern in accordance with the $M_{E}$ set to align IUI during 5-symbol extension is designed as
\begin{eqnarray}
(1,2,3,1,2), \textrm{ for user $[1,i]$},\,\,\,\,\, (1,1,1,2,2), \textrm{ for user $[2,i]$},
\end{eqnarray}
for $i \in \{1,2\}$. Consequently, IGI could be removed by the $m_{2}$ pattern in accordance with
$M_{G_{1}}=2$, and $M_{G_{2}}=2$. Then, $m_{2}$ pattern is
\begin{eqnarray}
(1,2,1), \textrm{ for user $[k,1]$},\,\,\,\,\, (1,1,2), \textrm{ for user $[k,2]$},
\end{eqnarray}
for $k \in \{1,2\}$. The preset mode pattern for each user is
\begin{eqnarray}
\mathbf{m}_{k,i}=\left\{\begin{array}{cc}
(1,2,3,1,2)\times(1,2,1), & \textrm{user $[1,1]$}, \\ (1,1,1,2,2)\times(1,2,1), & \textrm{user $[2,1]$}, \\
(1,2,3,1,2)\times(1,1,2), & \textrm{user $[1,2]$}, \\ (1,1,1,2,2)\times(1,1,2), & \textrm{user $[2,2]$}. \end{array} \right.
\label{eq:422}
\end{eqnarray}
The transmit signal from the transmitter is also designed successively according to the preset mode pattern.
The beamforming matrix for each user according to $m_{1}$ pattern during 5-symbol extension is
\begin{eqnarray}
\mathbf{V}^{1,1} \hspace{-2mm} &=& \hspace{-2mm} \left[\small \hspace{-1mm} \begin{array}{ccccc} (\mathbf{u}_{1}^{1,1})^{T} & (\mathbf{u}_{1}^{1,1})^{T} & (\mathbf{u}_{1}^{1,1})^{T} &
\mathbf{0}_{1\times6} & \mathbf{0}_{1\times6} \end{array} \hspace{-1mm} \right], \\ \nonumber
\mathbf{V}^{2,1} \hspace{-2mm} &=& \hspace{-2mm} \left[\small \hspace{-1mm} \begin{array}{ccccc} (\mathbf{u}_{1}^{2,1})^{T} & (\mathbf{u}_{2}^{2,1})^{T} & \mathbf{0}_{1\times4} &
(\mathbf{u}_{1}^{2,1})^{T} & (\mathbf{u}_{2}^{2,1})^{T} \end{array} \hspace{-1mm} \right], \\ \nonumber
\mathbf{V}^{1,2} \hspace{-2mm} &=& \hspace{-2mm} \left[\small \hspace{-1mm} \begin{array}{ccccc} (\mathbf{u}_{1}^{1,2})^{T} & (\mathbf{u}_{1}^{1,2})^{T} & (\mathbf{u}_{1}^{1,2})^{T} &
\mathbf{0}_{1\times6} & \mathbf{0}_{1\times6} \end{array} \hspace{-1mm} \right], \\ \nonumber
\mathbf{V}^{2,2} \hspace{-2mm} &=& \hspace{-2mm} \left[\small \hspace{-1mm} \begin{array}{ccccc} (\mathbf{u}_{1}^{2,2})^{T} & (\mathbf{u}_{2}^{2,2})^{T} & \mathbf{0}_{1\times4} &
(\mathbf{u}_{1}^{2,2})^{T} & (\mathbf{u}_{2}^{2,2})^{T} \hspace{-1mm} \end{array} \right],
\end{eqnarray}
where $\mathbf{u}^{k,i}\in \mathbb{C}^{M_{k,i} \times 1}$ is data streams intended to user $[k,i]$.
The transmit signal during 15-symbol extension according to the $m_{2}$ pattern is
\begin{eqnarray}
\mathbf{x}^{1,1} \hspace{-0.5mm} = \hspace{-0.5mm} \left[\mathbf{x}^{1,1}(1)^{T} \cdots \mathbf{x}^{1,1}(15)^{T} \right]^{T} \hspace{-0.5mm} = \hspace{-0.5mm} \left[\small \hspace{-1mm} \begin{array}{ccc} \mathbf{V}^{1,1} \hspace{-1mm} & \hspace{-1mm} \mathbf{V}^{1,1} \hspace{-1mm} & \hspace{-1mm} \mathbf{0}_{1\times 30} \end{array} \hspace{-1mm} \right]^{T}, \\ \nonumber
\mathbf{x}^{2,1} \hspace{-0.5mm} = \hspace{-0.5mm} \left[\mathbf{x}^{2,1}(1)^{T} \cdots \mathbf{x}^{2,1}(15)^{T} \right]^{T} \hspace{-0.5mm} = \hspace{-0.5mm} \left[\small \hspace{-1mm} \begin{array}{ccc} \mathbf{V}^{2,1} \hspace{-1mm} & \hspace{-1mm} \mathbf{V}^{2,1} \hspace{-1mm} & \hspace{-1mm} \mathbf{0}_{1\times 20} \end{array} \hspace{-1mm} \right]^{T}, \\ \nonumber
\mathbf{x}^{1,2} \hspace{-0.5mm} = \hspace{-0.5mm} \left[\mathbf{x}^{1,2}(1)^{T} \cdots \mathbf{x}^{1,2}(15)^{T} \right]^{T} \hspace{-0.5mm} = \hspace{-0.5mm} \left[\small \hspace{-1mm} \begin{array}{ccc} \mathbf{V}^{1,2} \hspace{-1mm} & \hspace{-1mm} \mathbf{0}_{1\times 30} \hspace{-1mm} & \hspace{-1mm} \mathbf{V}^{1,2} \end{array} \hspace{-1mm} \right]^{T}, \\ \nonumber
\mathbf{x}^{2,2} \hspace{-0.5mm} = \hspace{-0.5mm} \left[\mathbf{x}^{2,2}(1)^{T} \cdots \mathbf{x}^{2,2}(15)^{T} \right]^{T} \hspace{-0.5mm} = \hspace{-0.5mm} \left[\small \hspace{-1mm} \begin{array}{ccc} \mathbf{V}^{2,2} \hspace{-1mm} & \hspace{-1mm} \mathbf{0}_{1\times 20} \hspace{-1mm} & \hspace{-1mm} \mathbf{V}^{2,2} \end{array} \hspace{-1mm} \right]^{T}.
\label{eq:transmit_signal}
\end{eqnarray}
Each transmitter transmits $\mathbf{x}^{k,i}$, the received signal at receiver $[1,1]$ is (\ref{eq:received}).
\begin{figure*}[t]
\begin{eqnarray}
\label{eq:received}
\mathbf{y}_{1,1} \hspace{-3mm} &=& \hspace{-3mm}
\left[\small \begin{array}{c} y_{1,1}(1) \\ \vdots \\ y_{1,1}(15) \end{array} \right] =
\sum\limits_{^{\forall} k,i}
\left[\small \begin{array}{ccc} \mathbf{h}^{k,i}_{1,1}(l_{1,1}(1)) & \cdots & \mathbf{0}_{1 \times M_{k,i}} \\
\vdots & \ddots & \vdots \\
\mathbf{0}_{1 \times M_{k,i}} & \cdots & \mathbf{h}^{k,i}_{1,1}(l_{1,1}(15)) \end{array} \right]
\mathbf{x}^{k,i} \\ \nonumber
\hspace{-3mm} &=& \hspace{-3mm} \underbrace{\left[\small \begin{array}{c}
\mathbf{h}^{1,1}_{1,1}((1,1)) \\ \mathbf{h}^{1,1}_{1,1}((2,1)) \\ \mathbf{h}^{1,1}_{1,1}((3,1)) \\ \mathbf{0}_{1 \times 6} \\
\mathbf{0}_{1 \times 6} \\ \mathbf{h}^{1,1}_{1,1}((1,2)) \\ \mathbf{h}^{1,1}_{1,1}((2,2)) \\ \mathbf{h}^{1,1}_{1,1}((3,2)) \\
\mathbf{0}_{1 \times 6} \\ \mathbf{0}_{1 \times 6} \\ \mathbf{0}_{5 \times 6}
\end{array} \right]}_{\textrm{rank}=6}
\hspace{-1mm} \mathbf{u}^{1,1}_{1} \hspace{-1mm} + \hspace{-1mm}
\underbrace{\left[\small \begin{array}{cc}
\mathbf{h}^{2,1}_{1,1}((1,1)) & \mathbf{0}_{1 \times 4} \\ \mathbf{0}_{1 \times 4} & \mathbf{h}^{2,1}_{1,1}((2,1)) \\ \mathbf{0}_{1 \times 4} & \mathbf{0}_{1 \times 4} \\ \mathbf{h}^{2,1}_{1,1}((1,1)) & \mathbf{0}_{1 \times 4} \\
\mathbf{0}_{1 \times 4} & \mathbf{h}^{2,1}_{1,1}((2,1)) \\
\mathbf{h}^{2,2}_{1,1}((1,2)) & \mathbf{0}_{1 \times 4} \\ \mathbf{0}_{1 \times 4} & \mathbf{h}^{2,1}_{1,1}((2,2)) \\ \mathbf{0}_{1 \times 4} & \mathbf{0}_{1 \times 4} \\ \mathbf{h}^{2,1}_{1,1}((1,2)) & \mathbf{0}_{1 \times 4} \\
\mathbf{0}_{1 \times 4} & \mathbf{h}^{2,1}_{1,1}((2,2)) \\
\mathbf{0}_{5 \times 4} & \mathbf{0}_{5 \times 4}
\end{array} \right]}_{\textrm{rank}=4}
\hspace{-1mm} \left[\small \hspace{-1mm} \begin{array}{c} \mathbf{u}^{2,1}_{1} \\ \mathbf{u}^{2,1}_{2} \end{array} \hspace{-1mm} \right] \hspace{-1mm} + \hspace{-1mm}
\underbrace{\left[\small \begin{array}{c}
\mathbf{h}^{1,2}_{1,1}((1,1)) \\ \mathbf{h}^{1,2}_{1,1}((2,1)) \\ \mathbf{h}^{1,2}_{1,1}((3,1)) \\ \mathbf{0}_{1 \times 6} \\
\mathbf{0}_{1 \times 6} \\ \mathbf{0}_{5 \times 6} \\ \mathbf{h}^{1,2}_{1,1}((1,1)) \\
\mathbf{h}^{1,2}_{1,1}((2,1)) \\ \mathbf{h}^{1,2}_{1,1}((3,1))\\ \mathbf{0}_{1 \times 6} \\ \mathbf{0}_{1 \times 6}
\end{array} \right]}_{\textrm{rank}=3}
\hspace{-1mm} \mathbf{u}^{1,2}_{1} \hspace{-1mm} + \hspace{-1mm}
\underbrace{\left[\small \begin{array}{cc}
\mathbf{h}^{2,2}_{1,1}((1,1)) & \mathbf{0}_{1 \times 4} \\ \mathbf{0}_{1 \times 4} & \mathbf{h}^{2,2}_{1,1}((2,1)) \\ \mathbf{0}_{1 \times 4} & \mathbf{0}_{1 \times 4} \\ \mathbf{h}^{2,2}_{1,1}((1,1)) & \mathbf{0}_{1 \times 4} \\
\mathbf{0}_{1 \times 4} & \mathbf{h}^{2,2}_{1,1}((2,1)) \\
\mathbf{0}_{5 \times 4} & \mathbf{0}_{5 \times 4} \\
\mathbf{h}^{2,2}_{1,1}((1,1)) & \mathbf{0}_{1 \times 4} \\ \mathbf{0}_{1 \times 4} & \mathbf{h}^{2,2}_{1,1}((2,1)) \\ \mathbf{0}_{1 \times 4} & \mathbf{0}_{1 \times 4} \\ \mathbf{h}^{2,2}_{1,1}((1,1)) & \mathbf{0}_{1 \times 4} \\
\mathbf{0}_{1 \times 4} & \mathbf{h}^{2,2}_{1,1}((2,1))
\end{array} \right]}_{\textrm{rank}=2}
\hspace{-1mm} \left[\small \hspace{-1mm} \begin{array}{c} \mathbf{u}^{2,2}_{1} \\ \mathbf{u}^{2,2}_{2} \end{array} \hspace{-1mm} \right] \hspace{-1mm} ,
\end{eqnarray}
\hrulefill \vspace{-3mm}
\end{figure*}
It can be seen that the desired signal can be obtained by signal subtraction as
\begin{eqnarray}
\left[\small \hspace{-1mm} \begin{array}{c} \mathbf{h}^{1,1}_{1,1}((1,1)) \\ \mathbf{h}^{1,1}_{1,1}((2,1)) \\ \mathbf{h}^{1,1}_{1,1}((3,1)) \\ \mathbf{h}^{1,1}_{1,1}((1,2)) \\ \mathbf{h}^{1,1}_{1,1}((2,2)) \\ \mathbf{h}^{1,1}_{1,1}((3,2))
\end{array} \hspace{-1mm} \right] \hspace{-1mm} \mathbf{u}_{1}^{1,1} = \hspace{-1mm}
\left[\small \hspace{-1mm} \begin{array}{c} y_{1,1}(1)-y_{1,1}(4)-y_{1,1}(11)+y_{1,1}(14) \\
y_{1,1}(2)-y_{1,1}(5)-y_{1,1}(12)+y_{1,1}(15) \\ y_{1,1}(3)-y_{1,1}(13) \\ y_{1,1}(6)-y_{1,1}(9) \\ y_{1,1}(7)-y_{1,1}(10) \\
y_{1,1}(8) \end{array} \hspace{-1mm} \right] \hspace{-1mm} .
\end{eqnarray}
Other users also can obtain a desired signal by appropriate signal subtractions.
By this scheme, a total of 28/15 DoF is achievable during 15-symbol extension.
\section{Main Results}
\label{sec_Main}
In this section, we derive the variation of the achievable DoF and the symbol extension length of
grouping based supersymbol structure.

\textbf{Theorem 1: }
When $K$ users are grouped into $K_{G}$ groups ($K_{E}=K/K_{G}$) and
they have the number of preset mode groups as $M_{G_{i}}$ ($M_{E_{k}}=M_{k,i}/M_{G_{i}}$), respectively,
the achievable sum DoF is
\begin{eqnarray}
\label{eq:DoF}
\textrm{DoF}^{\textrm{grouping}}_{\textrm{sum}}=\frac{\sum\nolimits_{k=1}^{K_{E}}\frac{M_{E_{k}}}{M_{E_{k}}-1}}
{1+\sum\nolimits_{k=1}^{K_{E}}\frac{1}{M_{E_{k}}-1}} \times
\frac{\sum\nolimits_{i=1}^{K_{G}}\frac{M_{G_{i}}}{M_{G_{i}}-1}}{1+\sum\nolimits_{i=1}^{K_{G}}\frac{1}{M_{G_{i}}-1}}.
\end{eqnarray}
\begin{proof}
Let us prove the achievable sum DoF by calculating the individual DoF of user $[k,i]$ and adding it up.
According to \cite{IC}-\cite{IC2}, each user transmits $\prod\nolimits_{k'\neq k}(M_{k'}-1)$ symbols when $K$ users are
equipped with $M_{1},\ldots,M_{K}$ preset modes, respectively.
With the grouping result, user $[k,i]$ transmits $\prod\nolimits_{p\neq k}(M_{E_{p}}-1)\times
\prod\nolimits_{q\neq i}(M_{G_{q}}-1)$ symbols.
Then, the desired signal occupies
\begin{eqnarray}
\textrm{rk}_{\textrm{desired}} = M_{k,i}\prod\limits_{p\neq k}(M_{E_{p}}-1)
\prod\limits_{q\neq i}(M_{G_{q}}-1),
\end{eqnarray}
dimensions.
It is guaranteed that desired signals have separate subspaces from interference signals by the successive interference
alignment which satisfies condition (\ref{eq:con}).
Next, the interference signal from the user $[k',i]$ who are in the same group occupy $M_{G_{i}}$ dimensions
which are dropped from $M_{k',i}$ due to the IUI alignment by the $m_{1}$ pattern.
All the interference signals from intra-group users occupy
\begin{eqnarray}
\textrm{rk}_{\textrm{IUI}} = \sum\limits_{k' \neq k} \left( M_{G_{i}}
\prod\limits_{p'\neq k'}(M_{E_{p'}}-1)
\prod\limits_{q\neq i}(M_{G_{q}}-1)\right),
\end{eqnarray}
dimensions.
Lastly, interference signals from the inter-group user can be classified to two kinds, the interference signal
from the user $[k,i']$ who have the same $m_{1}$ pattern and that from the user $[k',i']$ who have no common
$m_{1}$ and $m_{2}$ pattern. The user $[k,i']$'s signal is aligned to $M_{E_{k}}$ dimensions, while the user $[k',i]$'s
signal is aligned to 1 dimension.
The dimension occupied by interference signals from inter-group users are
\begin{eqnarray}
\textrm{rk}_{\textrm{IGI}} = \sum\limits_{i' \neq i} \left( M_{E_{k}}
\prod\limits_{p\neq k}(M_{E_{p}}-1)
\prod\limits_{q'\neq i'}(M_{G_{q'}}-1)\right) \\ \nonumber
+ \sum\limits_{k' \neq k, i' \neq i}
\left( \prod\limits_{p'\neq k'}(M_{E_{p'}}-1)
\prod\limits_{q'\neq i'}(M_{G_{q'}}-1)\right).
\end{eqnarray}
Accordingly, the achievable DoF of user $[k,i]$ is
\begin{eqnarray}
\textrm{d}_{k,i} &=& \frac{\textrm{rk}_{\textrm{desired}}}{\textrm{rk}_{\textrm{desired}}+\textrm{rk}_{\textrm{IUI}}
+\textrm{rk}_{\textrm{IGI}}} \\ \nonumber
&=& \frac{\frac{M_{E_{k}}}{M_{E_{k}}-1}}
{1+\sum\nolimits_{p=1}^{K_{E}}\frac{1}{M_{E_{p}}-1}} \times
\frac{\frac{M_{G_{i}}}{M_{G_{i}}-1}}{1+\sum\nolimits_{q=1}^{K_{G}}\frac{1}{M_{G_{q}}-1}}.
\end{eqnarray}
Details are omitted due to space limit. Consequently, the achievable sum DoF is
\begin{eqnarray}
\textrm{DoF}^{\textrm{grouping}}_{\textrm{sum}}&=&\sum\limits_{i=1}^{K_{G}}\sum\limits_{k=1}^{K_{E}} d_{k,i} \\ \nonumber
&=&\frac{\sum\nolimits_{k=1}^{K_{E}}\frac{M_{E_{k}}}{M_{E_{k}}-1}}
{1+\sum\nolimits_{k=1}^{K_{E}}\frac{1}{M_{E_{k}}-1}} \times
\frac{\sum\nolimits_{i=1}^{K_{G}}\frac{M_{G_{i}}}{M_{G_{i}}-1}}{1+\sum\nolimits_{i=1}^{K_{G}}\frac{1}{M_{G_{i}}-1}}.
\end{eqnarray}
\end{proof}
\begin{remark}
The achievable sum DoF (\ref{eq:DoF}) can be interpreted as (the achievable DoF of the $m_{1}$ pattern with
$M_{E_{1}},\ldots,M_{E_{K_{E}}}$ preset modes) $\times$ (the achievable DoF of the $m_{2}$ pattern with
$M_{G_{1}},\ldots,M_{G_{K_{G}}}$ preset modes) since the achievable sum DoF for the MISO IC
introduced in \cite{IC}-\cite{IC2} is
\begin{eqnarray}
\label{eq:DoF_con}
\textrm{DoF}_{\textrm{sum}}=\frac{\sum\nolimits_{k=1}^{K}\frac{M_{k}}{M_{k}-1}}{1+\sum\nolimits_{k=1}^{K}\frac{1}{M_{k}-1}}.
\end{eqnarray}
\end{remark}
This result indicates that the grouping based supersymbol aligns interference signals in a hierarchical manner.
It means that IUI is aligned by $m_{1}$ pattern as if IGI does not exist,
and IGI is aligned by $m_{2}$ pattern with the assumption that IUI is perfectly removed.

\textbf{Theorem 2: }
The grouping based supersymbol structure reduces the symbol extension length as
\begin{eqnarray}
\mathcal{O}\left((\sqrt{M}-1)^{K-2\sqrt{K}}(\sqrt{M}+1)^{K}\right),
\end{eqnarray}
when we assume that $M_{E_{k}}=M_{E}=\sqrt{M}$, $M_{G_{i}}=M_{G}=\sqrt{M}$ for $k \in \{1,\ldots,K_{E}\}$, $i \in
\{1,\ldots,K_{G}\}$, $K_{E}=K_{G}=\sqrt{K}$, and $K \to \infty$.
\begin{proof}
Because the $m_{1}$ pattern repeats $m_{2}$ pattern length times, the symbol extension length of the grouping based
supersymbol structure ($\textrm{SL}^{\textrm{grouping}}$) is
\begin{eqnarray}
\label{eq:SL_grouping}
\textrm{SL}^{\textrm{grouping}} \hspace{-3mm} &=& \hspace{-3mm}
\left(\prod\limits_{k=1}^{K_{E}} (M_{E_{k}}-1)+
\sum\limits_{k=1}^{K_{E}}\prod\limits_{p=1,p\neq k}^{K_{E}} (M_{E_{p}}-1) \right) \\ \nonumber &\times\hspace{-3mm}&
\left(\prod\limits_{g=1}^{K_{G}} (M_{G_{g}}-1)+
\sum\limits_{g=1}^{K_{G}}\prod\limits_{q=1,q\neq g}^{K_{G}} (M_{G_{q}}-1) \right) \\ \nonumber
\hspace{-3mm} &=& \hspace{-3mm} \left((\sqrt{M}-1)^{\sqrt{K}}+\sqrt{K}(\sqrt{M}-1)^{\sqrt{K}-1}\right)^{2} \\ \nonumber
&\sim& \mathcal{O}(K(\sqrt{M}-1)^{2(\sqrt{K}-1)}).
\end{eqnarray}
As previously introduced, the supersymbol length without grouping is
\begin{eqnarray}
\label{eq:SL_con}
\textrm{SL} &=&
\prod\limits_{k=1}^{K}(M_{k}-1)+\sum\limits_{k=1}^{K}\prod\limits_{q=1,q\neq k}^{K}(M_{q}-1) \\ \nonumber
&=& (M-1)^{K}+K(M-1)^{K-1} \\ \nonumber
&\sim& \mathcal{O}(K(M-1)^{K-1}).
\end{eqnarray}
Therefore, the symbol extension length is reduced by the grouping based structure as
\begin{eqnarray}
\frac{\textrm{SL}}{\textrm{SL}^{\textrm{grouping}}}&\sim&\mathcal{O}\left(\frac{(M-1)^{K-1}}{(\sqrt{M}-1)^{2(\sqrt{K}-1)}}\right),\\ \nonumber
&\sim& \mathcal{O}\left((\sqrt{M}-1)^{K-2\sqrt{K}}(\sqrt{M}+1)^{K}\right).
\end{eqnarray}
\end{proof}
\begin{remark}
When $K_{G}=1$ and $M_{G_{1}}=1$, it is simply shown that the achievable DoF (\ref{eq:DoF}) and
supersymbol length (\ref{eq:SL_grouping}) of the grouping based
supersymbol structure coincide with those of the supersymbol with no grouping strategy,
(\ref{eq:DoF_con}) and (\ref{eq:SL_con}), respectively.
\end{remark}

\section{Simulations}
\label{sec_Sim}
We simulate our proposed scheme from the perspective of total achievable DoF with limited block length ($L$).
The length of the conventional supersymbol structure can be reduced by using fewer number of preset modes than already equipped
to meet a length constraint.
The grouping based supersymbol structure can be adjusted to $L$ by exploiting different number of user groups and preset mode groups. It also reduces the number of actually used preset modes as a conventional method to have a
symbol length not greater than $L$.

\begin{figure}[t]
    \centerline{\includegraphics[width=6.2cm, height=6.2cm]{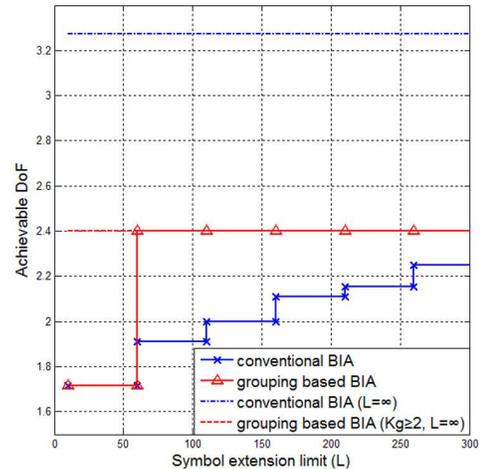}}
    \vspace{-2mm}
    \caption{Achievable DoF and maximum DoF when $(M,K)=(6,6)$.}
    \label{DoF_8}
    \vspace{-2mm}
\end{figure}

In Fig. \ref{DoF_8}, total achievable DoFs of conventional BIA and grouping based BIA with length constraint ($L$) are shown
when there are 6 users who have 6 preset modes, respectively. Moreover, it shows the maximum achievable DoF without length constraint ($L=\infty$). The maximum DoF of grouping based BIA is restricted to the case of $K_{G} \geq 2$ since we need to compare it to conventional BIA. It is observed that the grouping based BIA achieves larger DoF than the conventional BIA
with symbol extension limit, while the conventional BIA has larger maximum DoF than the grouping based BIA.
Although grouping based supersymbol with $K_{G} \geq 2$ cannot achieve larger maximum DoF, it has superiority
with given symbol extension limit due to its significant length reduction.

\begin{figure}[t]
    \centerline{\includegraphics[width=6.2cm, height=6.2cm]{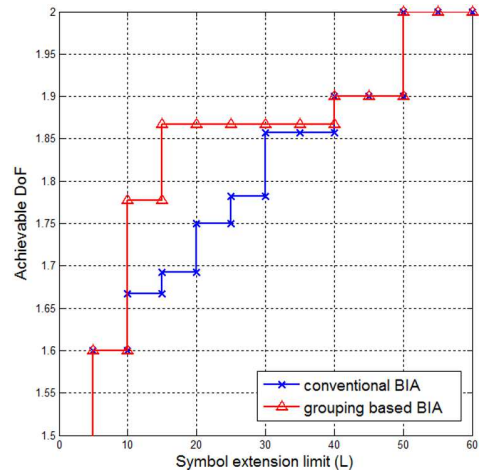}}
    \vspace{-2mm}
    \caption{Achievable DoF when $\mathcal{M}=\{6,6,4,4\}$.}
    \label{DoF_64}
    \vspace{-2mm}
\end{figure}

Fig. \ref{DoF_64} shows the case of 4-user IC when receivers have 6, 6, 4, and 4 preset modes, respectively.
It shows that the grouping based BIA achieves larger DoF than the conventional BIA when the length constraint is
$10 \leq L \leq 35$. For the relatively short symbol extension limit, the grouping based BIA adjusts $K_{G}=2$ to achieve
more DoF. When $L>35$, it sets $K_{G}=1$ since it can achieve larger DoF without grouping strategy, thereby its achievable
DoF is equal to that of conventional DoF.
It can be demonstrated that the grouping based BIA is a more efficient strategy with relatively short coherence time.

\section{Conclusion}
\label{sec_Con}
In this paper, we suggest the grouping based supersymbol structure for $K$-user MISO IC.
The grouping strategy reduces supersymbol length significantly with tolerable DoF loss.
The supersymbol structure is changed by the number of user and preset mode groups, thereby it can be adjusted to the
symbol extension limit to achieve larger DoF for given parameter.
The proposed BIA appears to be promising for practical channels with relatively short coherence time.
\section*{Acknowledgments}
\addcontentsline{toc}{section}{Acknowledgment}
This paper was supported in part by Samsung Electronics Co., Ltd, the National Research Foundation of Korea (NRF) grant
funded by the Korean Government (2013R1A1A2008956), and BK21 PLUS.



\end{document}